\documentclass[12pt]{article}

\usepackage[totalwidth=460truept,totalheight=622truept]{geometry}
\usepackage{latexsym,graphicx,amsfonts,amssymb,amsmath,xcolor,dsfont,cite}
\usepackage[hypertexnames=false,hidelinks]{hyperref}

\def\theequation{\arabic{section}.\arabic{equation}}
\newcommand{\sect}[1]{\setcounter{equation}{0}\section{#1}}
\renewcommand{\theequation}{\thesection.\arabic{equation}}
\global\arraycolsep=1truept
\renewcommand{\theequation}{\arabic{section}.\arabic{equation}}

\newcommand{\be}{\begin{equation}}
\newcommand{\ee}{\end{equation}}

\newcommand{\dd}{\mathrm{d}}

\usepackage[format=hang,font=small,textfont=it]{caption}

\begin{document}

\null

\vskip1truecm

\begin{center}
{\Large \textbf{Resummations In Higher-Curvature}}

\vskip.6truecm

{\Large \textbf{Modifications Of Starobinsky Inflation }}

\vskip1truecm

\textsl{{\large Marco Piva}}

\vskip .1truecm

{\textit{INFN, Sezione di Cagliari, Cittadella Universitaria, 09042 Monserrato, Italy}}

\vspace{0.2cm}

marco.piva@ca.infn.it
\vskip1truecm

\vskip .5truecm

\textbf{Abstract}
\end{center}
 We study higher-curvature corrections to Starobinsky inflation, which are expected to be generated at the quantum level. When treated perturbatively in the couplings $\gamma$ of the higher-dimensional operators, these terms produce inverse powers of the slow-roll parameter $\alpha$, casting doubt on the reliability of the slow-roll expansion. If instead $\gamma$ is kept exact, the expansion fails to reproduce the Starobinsky model smoothly as $\gamma\to 0$. We show that this problem is solved by resumming certain series in the ratio $\alpha/\gamma$. The resulting ``improved'' slow-roll expansion is regular in the limit $\gamma\to 0$ and gives the same result whether one expands first in $\gamma$ or in $\alpha$. We illustrate this explicitly for an $R^3$ term, obtain the power spectra, spectral indices and tensor-to-scalar ratio at next-to-leading order in the improved expansion, and show that results in the recent literature are recovered as the first-order Taylor expansion of ours. Comparing with the latest ACT, Planck and SPT data, we find that the resummed and first-order results differ appreciably only for values of $n_s$ outside the favored regions. Finally, we comment on the origin of higher-curvature operators in UV-complete theories, such as renormalizable quantum gravity, where the predictions of the Starobinsky model are stable.

\vfill\eject

\sect{Introduction}

Starobinsky inflation~\cite{Starobinsky:1980te} is one of the most successful models of the early universe. It is obtained by supplementing the Hilbert action with an $R^2$ term, and it makes sharp predictions for the spectral index and the tensor-to-scalar ratio. Yet it remains a nonrenormalizable theory. Quantum corrections inevitably generate operators of dimension higher than four, suppressed by powers of the mass scale of the theory. These are usually regarded as irrelevant at the inflationary scale, which is why the Starobinsky model is considered a good approximation.

This assumption has recently been reconsidered. The latest measurements of the Atacama Cosmology Telescope (ACT)~\cite{ACT:2025fju,ACT:2025tim} favor a slightly larger value of the scalar spectral index than the one predicted by the Starobinsky model, and several authors have investigated whether higher-curvature corrections can account for the shift~\cite{Modak:2025bjv,Ketov:2025cqg,Gialamas:2025ofz,Addazi:2025qra,Bezerra-Sobrinho:2025gfg,DelGrosso:2026zbg}. However, a consistent treatment of such corrections is not straightforward. Denote by $\gamma$ the dimensionless parameters of the higher-dimensional operators, collectively. If the calculations are performed perturbatively in $\gamma$, the inflationary observables acquire corrections proportional to inverse powers of the slow-roll parameter, and it is unclear whether the slow-roll expansion is still under control. If instead $\gamma$ is kept exact and one expands in slow-roll, the expansion works as usual but in the limit $\gamma\rightarrow 0$ the results do not reduce smoothly to those of the Starobinsky model. This was first noticed in~\cite{Piva:2025cqi}, where general classes of $f(R)$ theories were studied. In both cases the corrections cannot be regarded as genuine effective-field-theory (EFT) corrections to the Starobinsky model, since the limit in which they are switched off is either singular or not controlled.

The origin of the problem can be understood in the framework of~\cite{Piva:2025cqi}. There, $f(R)$ theories are classified according to the behavior of the beta function of the slow-roll parameter $\alpha$, which governs its evolution in conformal time, in close analogy with the running of a coupling constant in quantum field theory. Theories with a beta function that starts at $\mathcal{O}(\alpha^2)$, such as the Starobinsky model, are well behaved, since all the slow-roll parameters can be written as power series in $\alpha$. On the other hand, polynomial $f(R)$ theories of degree larger than two have a beta function that is linear in $\alpha$, and therefore fall outside this class. Since terms of this type are a subset of the possible local quantum corrections, it is important to understand whether a Starobinsky-like regime can be recovered in their presence.

In this paper we address this issue by following the renormalization-group (RG) strategy of~\cite{Piva:2025cqi,Anselmi:2021rye,Anselmi:2020shx}. We show that, at each order in the slow-roll expansion, certain series in the ratio $\alpha/\gamma$ can be resummed. The result is an ``improved'' slow-roll expansion, organized in powers of $\alpha$ with coefficients that are functions of $\gamma/\alpha$. In this expansion the beta function is quadratic in $\alpha$ up to functions of $\gamma/\alpha$, so the model effectively belongs to the well-behaved class and the results of~\cite{Piva:2025cqi} can be applied. The observables obtained in this way are smooth in $\gamma$, reproduce the Starobinsky predictions when $\gamma\rightarrow 0$, and recover the de Sitter solution that is lost in the naive expansion in $\gamma$. We also show that the final resummed result is the same whether one expands first in $\gamma$ and then in $\alpha$, or the other way round. We illustrate the procedure by extending the Starobinsky model with an $R^3$ term. The predictions at next-to-leading (NL) order depend on $\gamma$ through the combination $\sqrt{1 + \gamma^2/\alpha^2}$, and the formulas obtained in the recent literature are recovered as their first-order Taylor expansion. The same formulas hold for the opposite sign of $\gamma^2$, which is the relevant one for the ACT data, once the square root is replaced by $\sqrt{1 -\gamma^2/\alpha^2}$. This makes explicit the condition for the spectral index to be real, which is not visible in the truncated expansion.

We compare the resummed and the first-order results with the latest ACT-Planck and South Pole Telescope (SPT)~\cite{SPT-3G:2025bzu} data, and find that they differ appreciably only for values of $n_s$ outside the favored regions. We also point out that the combination with SPT data lowers the preferred value of $n_s$~\cite{Balkenhol:2025wms}, so that the Starobinsky model remains compatible with the data in this case, and with the ACT+Planck combination for a number of e-folds larger than 60. 

Finally, since the coefficients of higher-curvature operators should ultimately come from a UV-complete theory, we discuss the case of renormalizable quantum gravity with the fakeon prescription, where such operators arise only in a specific regime of the mass parameters. Indeed, the sensitivity of the observables to the $R^3$ term implies that the predictions of Starobinsky inflation, viewed as an effective theory, are exposed to unknown ultraviolet physics, which can shift $n_s$ by an amount comparable to the current uncertainty. In a perturbatively UV-complete theory this ambiguity is absent, and the predictions of the Starobinsky model are stable under quantum corrections, while the departures due to the fakeon prescription are confined mainly to the tensor-to-scalar ratio.

The paper is organized as follows. In Section 2 we review the classification of $f(R)$ theories and the derivation of the inflationary observables. In Section 3 we apply the resummation to the $R^3$ model, expanding both in $\alpha/\gamma$ and in $\gamma/\alpha$, and derive the predictions. In Section 4 we compare with other approaches in light of the ACT, Planck and SPT data. In Section 5 we comment on renormalizable gravity and on the stability of its predictions, and Section 6 contains our conclusions.

\sect{General treatment of $f(R)$ theories and inflationary observables}
In this section we review the procedure developed in~\cite{Piva:2025cqi} to classify $f(R)$ theories and derive the inflationary observables class by class.

We parametrize the action as
\be\label{eq:fRact}
S(g)=-\frac{1}{2\kappa^2}\int\sqrt{-g}f(R),\qquad \kappa^2=8\pi G,
\ee
where $G$ is the Newton constant, so that $M_{\mathrm{Pl}}=1/\kappa$ is the reduced Planck mass.

\subsection{Equations of motion and perturbative variable}

From~\eqref{eq:fRact} the equations of motion read
\be
F(R)R_{\mu\nu}-\frac{1}{2}g_{\mu\nu}f(R)+(g_{\mu\nu}\nabla^2-\nabla_{\mu}\nabla_{\nu})F(R)=0, \qquad F(R)\equiv\frac{\partial f}{\partial R},
\ee
which on a Friedmann-Lema\^itre-Robertson-Walker (FLRW) background
\be
g_{\mu\nu}^{\text{FLRW}}\mathrm{d}x^{\mu}\mathrm{d}x^{\nu}=\mathrm{d}t^2-a(t)^2\mathrm{d}x^{i}\mathrm{d}x^{i}
\ee
become
\be
H \dot{F}-F\dot{H}-FH^2-\frac{f}{6}=0,
\ee
\be
\ddot{F}+2\dot{H}F-H\dot{F}=0,
\ee
where $H\equiv\dot{a}/a$ is the Hubble parameter. In our conventions $R=-6(\dot H+2H^2)$, which is negative during inflation.

It is important to notice that, in single-field inflation, one slow-roll parameter is enough to write every quantity as a perturbative series in it, including the additional parameters that are often used in the literature. We find that using a single parameter makes the derivation clearer and less prone to errors, since different slow-roll parameters might otherwise contribute at different orders. Therefore, we choose one slow-roll parameter $\alpha$ and write everything else as a power series in $\alpha$. In~\cite{Piva:2025cqi} it was shown that a particularly useful choice is
\be\label{eq:coupling}
\alpha\equiv-\frac{\dot{F}}{2HF}.
\ee
We also define various useful quantities
\be\label{eq:epsvbeta}
\varepsilon\equiv -\frac{\dot{H}}{H^2}, \qquad v\equiv-\frac{1}{Ha\tau}, \qquad \beta_{\alpha}\equiv\frac{\mathrm{d}\alpha}{\mathrm{d}\ln|\tau|},
\ee
where $\tau$ is the conformal time. All of them can be shown to be power series in $\alpha$. In particular, the \emph{beta function} $\beta_{\alpha}$ is an important quantity used to classify various models. In terms of~\eqref{eq:epsvbeta} the equations of motion turn into
\be\label{eq:betaeom}
-v\beta_{\alpha}=\alpha(1+2\alpha)-(1-\alpha^2)\varepsilon,
\ee
\be\label{eq:varepsilon}
\varepsilon=1+2\alpha+\frac{f}{6FH^2}.
\ee

\subsection{Slow-roll parameters}

It is easy to show that if $\beta_{\alpha}$ is at least linear in $\alpha$ and with a negative coefficient, then $\alpha$ as a function of $\tau$ goes to zero in the infinite past, and so does $\varepsilon$. Therefore, the background spacetime approaches the de Sitter one in the same limit, which is a necessary condition for inflation to happen. However, the linearity in $\alpha$ does not guarantee that the other slow-roll parameters also vanish in the infinite past. For example, the parameter
\be
\varepsilon_2\equiv\frac{\dot{\varepsilon}}{\varepsilon H}
\ee
turns out to be a power series in $\alpha$ starting from a constant~\cite{Piva:2025cqi} (see also below), while for quadratic $\beta_{\alpha}$ we have that $\varepsilon_2=\mathcal{O}(\alpha)$.
More generally, we can define the slow-roll parameters
\be
\varepsilon_n\equiv\frac{\dot{\varepsilon}_{n-1}}{H\varepsilon_{n-1}}, \qquad \varepsilon_1\equiv\varepsilon
\ee
and show that they are all at least $\mathcal{O}(\alpha)$ if the beta function is at least $\mathcal{O}(\alpha^2)$. To do so, we proceed as follows. First, from~\eqref{eq:betaeom} we have
\be\label{eq:eps1}
\varepsilon=\frac{\alpha+2\alpha^2+v\beta_{\alpha}}{1-\alpha^2},
\ee
which, using that\footnote{Since $v=1$ in de Sitter spacetime we can write $v=1+\mathcal{O}(\alpha)$.}
\be
v=1+\mathcal{O}(\alpha), \qquad \beta_{\alpha}=\mathcal{O}(\alpha^m),
\ee
gives
\be\label{eq:epsalpha}
\varepsilon=\alpha+\mathcal{O}(\alpha^2).
\ee
Now, to prove that all the $\varepsilon_i$ are at least of order $\alpha$ we first prove by induction that they are $\mathcal{O}(\varepsilon)$, so that, from~\eqref{eq:epsalpha}, they are also $\mathcal{O}(\alpha)$. Indeed, using $\beta_{\varepsilon}=\mathcal{O}(\varepsilon^m)$ with $m\geq2$, defined analogously to~\eqref{eq:epsvbeta}, we have
\begin{equation}
\varepsilon_2=-\frac{v\beta_{\varepsilon}}{\varepsilon}=\mathcal{O}(\varepsilon^{m-1}), \qquad \varepsilon_{3}=-\frac{v\beta_{\varepsilon}}{\varepsilon_2}\frac{\partial\varepsilon_2}{\partial\varepsilon}=\mathcal{O}(\varepsilon^{m-1})
\end{equation}
and so on. Assuming that the result holds up to a fixed $n$, it is straightforward to show that it holds also for $n+1$, which proves the statement.

To summarize, we have translated the typical slow-roll requirements into conditions on $\alpha$ and $\beta_{\alpha}$, i.e.
\be
|\varepsilon_i|\ll 1 \ \  \forall i \qquad \Longleftrightarrow \qquad \alpha\ll 1 \quad \land \quad \beta_{\alpha}=\mathcal{O}(\alpha^m), \ \ m\geq 2.
\ee

However, as shown in~\cite{Piva:2025cqi}, polynomial $f(R)$ theories of degree larger than two have a linear beta function. Nevertheless, such theories are expected to describe small corrections to the Starobinsky model, at least in some regime, and terms of this type are only a subset of the possible local quantum corrections. Therefore, it is important to investigate whether a mechanism to recover a Starobinsky-like inflation exists in their presence.

We call \emph{class I} the set of theories with a quadratic beta function, \emph{class II} those with a cubic one, and so on. Theories with a linear beta function, such as polynomial $f(R)$ of degree larger than two, form a special \emph{class 0}.

\subsection{Running slow-roll parameter}

As in~\cite{Piva:2025cqi}, the definition of the beta function turns the evolution of $\alpha$ into a first-order differential equation. We introduce an arbitrary scale $k$ of mass dimension one and define the \emph{rescaled conformal time}
\be
\eta\equiv-k\tau,
\ee
together with its logarithm
\be
\ell\equiv \ln\eta.
\ee
Within the RG analogy developed in~\cite{Anselmi:2020shx,Anselmi:2021rye}, $\eta$ plays the role of the renormalization scale. The evolution equation for $\alpha$ then reads
\be\label{eq:eqalpha}
\frac{\mathrm{d}\alpha(\ell)}{\mathrm{d}\ell}=\beta_{\alpha}(\alpha),
\ee
and its solution defines a \emph{running slow-roll parameter}.

We solve it with the boundary condition $\alpha(\ell=0)=\alpha_k$, i.e.\ at $\tau=-1/k$. Later on, $k$ will be identified with the modulus of the spatial momentum of the perturbations, so that, once $\alpha$ is known at a pivot scale $k_*$, it can be determined at any other scale. As in the standard RG, this follows from the fact that $k$ is introduced by hand and cannot affect $\alpha$, which gives
\be\label{eq:alphakeq}
\frac{\mathrm{d}\alpha}{\mathrm{d}\ln k}=0, \qquad \Rightarrow \qquad \frac{\mathrm{d}\alpha_k}{\mathrm{d}\ln k}=-\beta_{\alpha}(\alpha_k).
\ee
Equation~\eqref{eq:eqalpha} can be solved by truncating the beta function. Since the formal expressions do not depend on the frame, we can take them from~\cite{Anselmi:2021rye}. For classes I and II, at leading-log order (LLO), we have
\be\label{eq:alpharun12}
\alpha^{\text{I}}(\ell)=\frac{\alpha_k}{1-b_0\alpha_k\ell}, \qquad \alpha^{\text{II}}(\ell)=\frac{\alpha_k}{\sqrt{1-2b_1\alpha_k^2\ell}}.
\ee
We omit the next-to-leading-log (NLL) corrections, because here we can rely directly on the results of~\cite{Anselmi:2021rye}. What matters is that power spectra and spectral indices are eventually expressed as expansions in $\alpha_k$, whose value at the pivot scale $k_*$ is fixed by the data, while the second equation in~\eqref{eq:alphakeq} gives $\alpha_k$ at any other scale. For instance, in class I
\be
\alpha_k=\frac{\alpha_{k_*}}{1+b_0\alpha_{k_*}\ln\left(k/k_*\right)}.
\ee
As a consequence, our results contain the leading-log corrections already resummed in the running parameter, whereas standard approaches obtain them with considerably more effort.

The last ingredient of the RG analogy is a Callan-Symanzik-type equation for correlators, which here expresses the conservation of the power spectra $\mathcal{P}$ on superhorizon scales, $\eta\rightarrow 0$
\be
\frac{\mathrm{d}\mathcal{P}}{\mathrm{d}\ln|\tau|}=0.
\ee
Viewing $\mathcal{P}$ as a function of $\alpha(\ell)$ and $\alpha_k$, this conservation implies that the dependence on $\alpha(\ell)$ drops out, so that $k$ enters only through $\alpha_k$. Further details are given in~\cite{Anselmi:2020shx}.

\subsection{Predictions}
To derive the predictions, we trade every time derivative for a derivative with respect to $\alpha$ through the relation
\be\label{eq:dt}
\frac{\mathrm{d}}{\mathrm{d}t}=-vH\beta_{\alpha}\frac{\mathrm{d}}{\mathrm{d}\alpha}.
\ee
This allows us to express $H$, $v$ and $F$ as expansions in $\alpha$, by solving perturbatively the equations
\be\label{eq:vFHeq}
\beta_{\alpha}\frac{\partial v}{\partial\alpha}=1-v-\varepsilon, \qquad v\beta_{\alpha}\frac{\partial F}{\partial\alpha}=2\alpha F, \qquad v\beta_{\alpha}\frac{\partial H}{\partial\alpha}= H\varepsilon,
\ee
which follow directly from the definitions of $v$, $\alpha$ and $\varepsilon$, respectively. Depending on the class, the solutions may contain overall nonpolynomial factors.

Cosmological perturbations obey a Mukhanov-Sasaki equation of the form
\be\label{eq:MSeq}
w''+w-\frac{2+\sigma}{\eta^2}w=0,
\ee
where $\sigma$ is a power series in $\alpha$ that depends on the model and on the type of perturbation. For tensor and scalar perturbations it is given, respectively, by
\be\label{eq:sigmat}
\sigma_{\text{t}}=\frac{2(1-v^2)-\alpha(4+\alpha)}{v^2},
\ee
\be\label{eq:sigmas}
\sigma_{\text{s}}=\sigma_{\text{t}}-\frac{\beta_{\alpha}}{ \alpha}\left[\frac{2}{v}+\frac{1}{1-\alpha}\left(1-\frac{\mathrm{d}\beta_{\alpha}}{\mathrm{d}\alpha}-\frac{2\beta_{\alpha}}{1-\alpha}\right)\right].
\ee

The solution can be obtained as an expansion in powers of $\alpha_k$ and then used to compute the power spectra. It can be decomposed as~\cite{Piva:2025cqi,Anselmi:2020shx}
\be
\eta w(\eta)=Q(\ln\eta)+Y(\eta),
\ee
where $Y$ vanishes in the superhorizon limit $\eta\rightarrow 0$. Therefore, in that limit the power spectra depend only on $Q$, and in particular on $Q_0\equiv Q(0)$, i.e.\ $Q$ evaluated at $\ell=0$, which is a power series in $\alpha_k$. For a generic $f(R)$ theory the power spectra read~\cite{Piva:2025cqi,Anselmi:2020shx}
\be\label{eq:generalPtPs}
 \mathcal{P}_{\text{t}}=\frac{4\kappa^2v(\alpha_k)^2H(\alpha_k)^2\left|Q_{0\text{t}}(\alpha_k)\right|^2}{\pi^2F(\alpha_k)}, \qquad \mathcal{P}_{\text{s}}=\frac{\kappa^2v(\alpha_k)^2H(\alpha_k)^2(1-\alpha_k)^2\left|Q_{0\text{s}}(\alpha_k)\right|^2}{12\pi^2\alpha_k^2F(\alpha_k)}.
\ee

From these expressions, using~\eqref{eq:alphakeq} and~\eqref{eq:vFHeq}, we obtain the tensor-to-scalar ratio and the spectral indices
\be\label{eq:generalr}
r=48\left(\frac{\alpha_k}{1-\alpha_k}\right)^2\left|\frac{Q_{0\text{t}}(\alpha_k)}{Q_{0\text{s}}(\alpha_k)}\right|^2,
\ee

\be\label{eq:generalnt}
n_{\text{t}}=2\left[1-\frac{1-\alpha_k}{v(\alpha_k)}-\beta_{\alpha}(\alpha_k)\frac{\mathrm{d}\ln|Q_{0\text{t}}(\alpha_k)|}{\mathrm{d}\alpha_k}\right],
\ee

\be\label{eq:generalns}
n_{\text{s}}-1=2\left\{1-\frac{1-\alpha_k}{v(\alpha_k)}+\beta_{\alpha}(\alpha_k)\left[\frac{1}{\alpha_k(1-\alpha_k)}-\frac{\mathrm{d}\ln|Q_{0\text{s}}(\alpha_k)|}{\mathrm{d}\alpha_k}\right]\right\}.
\ee

These formulas reduce the computation of the inflationary observables of any $f(R)$ theory to the determination of the beta function and of the expansions of $v$, $F$, $H$ and $Q_0$ in powers of $\alpha_k$. In the next section we apply this procedure to the Starobinsky model extended with an $R^3$ term, which belongs to class 0, and show how a resummation of certain series in $\alpha_k/\gamma$ restores a well-behaved expansion.

\sect{Starobinsky inflation with a $R^3$ term}

We consider the Starobinsky model extended with a cubic term,
\be\label{eq:R3act}
S(g)=-\frac{1}{2\kappa^2}\int\sqrt{-g}\left[R-\frac{1}{6M^2}R^2+\frac{\gamma^2}{16M^4}R^3\right], \qquad \kappa^2=8\pi G,
\ee
where $M$ is the inflaton mass and $\gamma$ is a dimensionless parameter. We choose a positive sign for the coefficient of $R^3$ because it allows us to expand in both $\alpha$ and $\gamma$, in either order. The replacement $\gamma^2\rightarrow-\gamma^2$ is discussed later.

The background equation of motion in terms of $\varepsilon$ is
\be\label{eq:eomR3}
\left[1+\frac{27H^2\gamma^2}{2M^2}\left(1-\frac{\varepsilon}{2}\right)\right]\frac{\dot{\varepsilon}}{H}=-3\varepsilon\left(1-\frac{\varepsilon}{2}\right)+\frac{M^2}{2 H^2}-\frac{9\gamma^2H^2}{8M^2}\left(2-\varepsilon\right)^2\left(1+10\varepsilon\right).
\ee
The theory~\eqref{eq:R3act} can be studied in two different ways. We either assume from the start that $\gamma$ is small and expand in it first, and then perform the slow-roll expansion, or leave the exact dependence on $\gamma$ and expand in $\alpha$ first. In the next subsection we show both possibilities, starting from the latter. As we show below in both cases we have to deal with inverse powers, either of $\alpha$ or of $\gamma$, which are both unsatisfactory. In fact, inverse powers of $\alpha$ make the slow-roll expansion unreliable, while inverse powers of $\gamma$ do not allow for a smooth transition between Starobinsky model and its extensions, which in turn cannot be regarded as EFT corrections. This is resolved by means of perturbative resummations, which allows us to treat the ratio $\alpha/\gamma$ as of order one. 

It is also worth noting that, when $\gamma$ is treated exactly, de Sitter spacetime is an exact solution of the equation~\eqref{eq:eomR3}. Indeed, if we set $H=\bar{H}=\text{constant}$, equation~\eqref{eq:eomR3} reduces to
\be
\frac{M^2}{2 \bar{H}^2}-\frac{9\gamma^2\bar{H}^2}{2M^2}=0,
\ee
which has a real and positive solution
\be\label{eq:desitterH}
\bar{H}=\frac{M}{\sqrt{3\gamma}}.
\ee
This solution is nonperturbative in $\gamma$ and is lost if we expand in $\gamma$ from the beginning. However, it is recovered (although only in the case $\gamma^2>0$) once the resummations mentioned above are performed (see below).

\subsection{Expansion in $\alpha/\gamma$}
The fact that de Sitter spacetime is an exact solution allows us to write a generic ansatz for the expansion of $H$. Since in de Sitter spacetime $\alpha=0$, we can write  
\be
H(\alpha)=\bar{H}\sum_{n=0}^{\infty}h_n\alpha^n, \quad h_0=1,
\ee
where $h_i$ are $\gamma$-dependent constant coefficients. In order to write all the quantities as expansions in $\alpha$ it is more convenient to write $H$ and $\varepsilon$ as functions of $R$ and use the third equation in~\eqref{eq:vFHeq} to find the expansions for $R$, which is also a power series 
\be
R(\alpha)=\bar{R}\sum_{n=0}^{\infty}r_n\alpha^n, \quad r_0=1,
\ee
since
\be\label{eq:RHeps}
R=-12H^2\left(1-\frac{\varepsilon}{2}\right).
\ee  
The constant $\bar{R}$ can be easily found from~\eqref{eq:RHeps}, which gives
\be
\bar{R}=-\frac{4M^2}{\gamma}.
\ee
Combining~\eqref{eq:RHeps} with~\eqref{eq:varepsilon} we get
\be
H=-\frac{R\sqrt{4M^2-\gamma^2R}}{3\sqrt{(1-2\alpha)\left(48M^4-16M^2R+9\gamma^2R^2\right)}}.
\ee
Similarly, we can write $\dot{\alpha}$ in terms of $R$ and $\alpha$ itself by means of the equation of motion
\be\label{eq:alphadot}
\frac{\dot{\alpha}}{H}=\alpha(1+2\alpha)-(1-\alpha^2)\varepsilon.
\ee
Finally, we plug everything into the third equation in~\eqref{eq:vFHeq}, which can be written as
\be\label{eq:Heqalpha}
\dot{\alpha}\frac{\partial H}{\partial\alpha}=\varepsilon H^2
\ee
and perturbatively solve it for the coefficient $r_i$. These coefficients are rather cumbersome functions of $\gamma$ and we do not report them here. We find that they can be written as
\be
r_i=\frac{\tilde{r}_i}{\gamma^i},
\ee
where $\tilde{r}_i$ are regular in $\gamma=0$. This suggests that the expansion for $R$ might contain an expansion in powers of $\alpha/\gamma$ that could be resummed. Indeed, if we expand the coefficients $r_i$ in $\gamma$ the series for $R$ can be reorganized as
\be\label{eq:Rxi}
R=\bar{R}\sum_{n=0}^{\infty}R_n(\xi)\gamma^n,
\ee
where we have defined the variable
\be
\xi\equiv\frac{\alpha}{\gamma}
\ee
and $R_i(\xi)$ are power series. For example, we have
\be
R_0(\xi)=1-\xi +\frac{\xi ^2}{2}-\frac{\xi ^4}{8}+\frac{\xi ^6}{16}-\frac{5 \xi ^8}{128}+\frac{7 \xi ^{10}}{256}+\mathcal{O}(\xi^{12}),
\ee
which is the Taylor expansion of 
\be
\frac{1}{\xi +\sqrt{1+\xi ^2}}
\ee
around $\xi=0$. This means that at the leading order in the expansion~\eqref{eq:Rxi} we have
\be
R\simeq -\frac{4M^2}{\alpha+\sqrt{\alpha^2+\gamma^2}},
\ee
which is now regular in $\gamma=0$ and correctly reproduces the expression for $R$ in that limit. All functions $R_i(\xi)$ can be found by inserting~\eqref{eq:Rxi} into~\eqref{eq:Heqalpha} and solve it perturbatively in $\gamma$. Order by order the resulting set of equations are algebraic in $R_i(\xi)$ and can be easily solved. We find
\be
R_0(\xi)=\frac{1}{\xi +\sqrt{1+\xi ^2}}, \qquad R_1(\xi)=\frac{\xi  \left(6 \xi -11 \sqrt{1+\xi ^2}\right)}{3+3\xi  \left(\xi +\sqrt{1+\xi ^2}\right)},
\ee

\be
R_2(\xi)=\frac{1}{9} \xi  \left(13-218 \xi ^2\right)+\frac{\xi ^2 \left(155+627 \xi ^2+436 \xi ^4\right)}{18 \left(1+\xi
   ^2\right)^{3/2}},
\ee

\be
R_3(\xi)=\frac{1}{27} \xi ^2 \left(488+4019 \xi ^2\right)-\frac{\xi  \left(419+4764 \xi ^2+17599 \xi ^4+21076 \xi ^6+8038
   \xi ^8\right)}{54 \left(1+\xi ^2\right)^{5/2}},
\ee
and so on. Once multiplied by the appropriate factor $\gamma^{n-1}$, all these functions become regular in $\gamma=0$. In particular, they reproduce the correct expansion in $\alpha$ of Starobinsky model
\be
\left.R(\alpha)\right|_{\gamma=0}=-\frac{2M^2}{\alpha}\left[1-\frac{5 \alpha }{3}-\frac{\alpha ^2}{9}-\frac{5 \alpha ^3}{27}-\frac{4 \alpha ^4}{27}-\frac{40 \alpha
   ^5}{81}+\frac{16 \alpha ^6}{81}+\mathcal{O}(\alpha^7)\right].
\ee
Following the same procedure we can find also the expansions for $H$ and $v$ reorganized as
\be
H=\bar{H}\sum_{n=0}^{\infty}H_n(\xi)\gamma^n, \qquad v=1+\sum_{n=1}^{\infty}V_n(\xi)\gamma^n.
\ee
We obtain
\be
H_0(\xi)=\frac{1}{\sqrt{\xi+\sqrt{1+\xi^2}}}, \qquad H_1(\xi)=-\frac{\xi  \left[19+7 \xi  \left(\xi +\sqrt{1+\xi ^2}\right)\right]}{12 \sqrt{1+\xi ^2} \left(\xi +\sqrt{1+\xi
   ^2}\right)^{3/2}},
\ee
\be
V_1(\xi)=-\xi, \qquad V_2(\xi)=-3\xi^2, \qquad V_3(\xi)=-3 \xi ^2 \left(\xi +4 \sqrt{1+\xi ^2}\right).
\ee
Again, after the resummations, the expansions for $H$ and $v$ are regular in $\gamma=0$ and correctly reproduce the expansions in the Starobinsky case~\cite{Piva:2025cqi}
\be
\left.H(\alpha)\right|_{\gamma=0}=\frac{M}{\sqrt{6\alpha}}\left[1-\frac{7 \alpha }{12}-\frac{77 \alpha ^2}{288}-\frac{883 \alpha ^3}{3456}-\frac{50045 \alpha
   ^4}{165888}-\frac{311731 \alpha ^5}{663552}+\mathcal{O}(\alpha^6)\right],
\ee
\be
\left.v(\alpha)\right|_{\gamma=0}=1-\alpha -3 \alpha ^2-15 \alpha ^3-109 \alpha ^4-\frac{3103 \alpha ^5}{3}+\mathcal{O}(\alpha^6).
\ee
The functions $R_i(\xi)$, $H_i(\xi)$ and $V_i(\xi)$ ha been derived up to $i=6$. 

To carry out the calculations in the slow-roll expansion, we need to reorganize our results. So far we have expressed every quantity as an expansion in powers of $\gamma$, with coefficients that are functions of the ratio $\xi$, which we treat as of order one. However, it is more convenient to have an expansion in powers of $\alpha$, still with coefficients that are functions of $\xi$, in close analogy with the resummation of running couplings in quantum field theory. It is easy to rearrange the expansions in such a way, which we call \emph{improved} slow-roll expansion. Indeed, we can just define the ratio
\be
\zeta\equiv\frac{\gamma}{\alpha}=\frac{1}{\xi}
\ee
and rewrite everything in terms of $\alpha$ and $\zeta$. For example, for $R$ we have
\be
R=-\frac{4M^2}{\alpha\zeta}\sum_{n=0}^{\infty}R_n(1/\zeta)\zeta^n\alpha^n=-\frac{2M^2}{\alpha}\sum_{n=0}^{\infty}\tilde{R}_n(\zeta)\alpha^n,
\ee
where we have defined
\be\label{eq:Rtilden}
\tilde{R}_n(\zeta)\equiv2\zeta^{n-1}R_n(1/\zeta).
\ee
The expansions for $H$ and $v$ can be rearranged in the same way. As a result, neither inverse powers of $\gamma$ nor of $\alpha$ appear in the expansion, since they are absorbed in the resummation, except for an overall inverse power of $\alpha$, which is harmless. In the next subsection we show that the same improved expansion is obtained by starting from the opposite order, i.e. by expanding in $\gamma$ first and then in $\alpha$.

Using this arrangement we can derive $\beta_{\alpha}$ from its definition and~\eqref{eq:alphadot}
\be
\beta_{\alpha}=-2 \alpha ^2 \sqrt{1+\zeta ^2}+\alpha ^3 \left(\frac{31}{3}\left(1+\zeta^2\right)+\frac{4}{\sqrt{1+\zeta ^2}}-16
   \sqrt{1+\zeta ^2}\right)+\mathcal{O}_4,
\ee
where $\mathcal{O}_n$ denotes $\mathcal{O}(\alpha^n)$, with functions of $\zeta$ counted as $\mathcal{O}(\alpha^0)$.
In these variables it is also easier to read the limit $\gamma=0$, which corresponds to $\zeta=0$ and gives the beta function of Starobinsky model~\cite{Piva:2025cqi}
\be
\left.\beta_{\alpha}\right|_{\gamma=0}=-2 \alpha ^2-\frac{5 \alpha ^3}{3}-\frac{65 \alpha ^4}{9}-\frac{1111 \alpha ^5}{27}+\mathcal{O}(\alpha^6).
\ee

After these resummations, the theory~\eqref{eq:R3act} belongs to a sort of ``improved'' class I, since its beta function is quadratic in $\alpha$ up to functions of $\zeta$. Therefore, we can use the formulas for the power spectra and the spectral indices of class I derived in~\cite{Piva:2025cqi}, with the coefficients of the beta function promoted to functions of $\zeta$. To make sure that no subtlety invalidates this procedure, we have also repeated the derivation directly with the new expansions, and the two results coincide.

\subsection{Expansion in $\gamma/\alpha$}

The theory~\eqref{eq:R3act} can also be studied by treating $\gamma$ as small from the very beginning. For example,~\eqref{eq:eomR3} can be solved perturbatively in $\gamma$, which gives the Starobinsky result plus corrections that are polynomial in $\gamma$. Proceeding in this way, we can repeat the calculations of the previous subsection. Then, all the formulas reduce smoothly to those of the Starobinsky model for $\gamma\rightarrow0$, at the price of containing inverse powers of $\alpha$, which potentially jeopardize the slow-roll expansion. In this case the de Sitter solution~\eqref{eq:desitterH} is lost, being nonperturbative in $\gamma$, but it is recovered after the resummation.

Expanding in $\gamma$ first and following the same steps of the previous subsection leads us to parametrize the expansion of $R$ as
\be\label{eq:Rexpansion}
R(\alpha)=-\frac{2M^2}{\alpha}\sum_{n,m=0}^{\infty}r_{nm}\alpha^n\left(\frac{\gamma}{\alpha}\right)^{2m},
\ee
where the $r_{nm}$ are numerical coefficients, independent of $\gamma$, which can be found by solving~\eqref{eq:Heqalpha} perturbatively. Rewriting~\eqref{eq:Rexpansion} in terms of $\zeta$ we have
\be
R(\alpha)=-\frac{2M^2}{\alpha}\sum_{n=0}^{\infty}\hat{R}_n(\zeta)\alpha^n,
\ee
where the $\hat{R}_n(\zeta)$ are the power series
\be
\hat{R}_n(\zeta)=\sum_{m=0}^{\infty}r_{nm}\zeta^{2m},
\ee
which are the expansions of~\eqref{eq:Rtilden} around $\zeta=0$. Therefore,
\be
\hat{R}_n(\zeta)=\tilde{R}_n(\zeta), \qquad \forall n\geq 0.
\ee
This equality means that the series $\hat{R}_n(\zeta)$ converges to the function $\tilde{R}_n(\zeta)$ for $|\zeta|<1$, the radius of convergence being set by the branch points of $\tilde{R}_n$ at $\zeta=\pm i$. Beyond it, $\tilde{R}_n(\zeta)$ provides the analytic continuation. The same holds for all the other expansions.

To summarize, the final resummation is the same whether we expand in $\gamma$ first and then in $\alpha$, or vice versa.

\subsection{Predictions}
Before proceeding with the application of formulas~\eqref{eq:generalPtPs} we solve the equation for the improved beta function at the leading order
\be
\frac{\dd\alpha(\ell)}{\dd\ell}=-2\alpha\sqrt{\alpha^2+\gamma^2}, \qquad \alpha(0)=\alpha_k,
\ee
\be
\alpha(\ell)=\frac{2 e^{2 \ell \gamma } \gamma  \alpha _k \sqrt{\alpha _k^2-2 \gamma  \left(\sqrt{\gamma ^2+\alpha _k^2}-\gamma
   \right)}}{\left(e^{4 \ell \gamma }-1\right) \alpha _k^2+2 \gamma  \left(\sqrt{\gamma ^2+\alpha _k^2}-\gamma
   \right)}.
\ee
If we expand in $\alpha_k$ while assuming the ratio $\zeta_k\equiv\gamma/\alpha_k$ of order one, we get a more familiar form~\cite{Anselmi:2021rye}
\be
\alpha(\ell)=\alpha _k-2 \ell \alpha _k^2 \sqrt{1+\zeta _k^2}+2 \ell^2 \alpha _k^3 \left(2+\zeta _k^2\right)+\mathcal{O}_3, 
\ee
where now $\mathcal{O}_n$ means $\mathcal{O}(\alpha_k^n)$ up to functions of $\zeta_k$. This expansion can be used as usual to obtain the power spectra from formulas~\eqref{eq:generalPtPs}. As mentioned above, we also solved the Mukhanov-Sasaki equation with the improved expansions and the result is the same as in~\cite{Piva:2025cqi} with the formal substitution of the beta function coefficient with $\zeta_k$-dependent ones. 

Then, the power spectra at the NL order read
\be\label{eq:Ptimproved}
\mathcal{P}_{\text{t}}=\frac{M^2\kappa^2}{2\pi^2}\left[1-3\alpha_k\sqrt{1+\zeta_k^2}+\mathcal{O}_2\right],
\ee

\be
\mathcal{P}_{\text{s}}=\frac{M^2\kappa^2}{96\pi^2\alpha_k^2}\left\{1-\alpha _k \left[2-\left(5-4 c_E\right) \sqrt{1+\zeta _k^2}\right]+\mathcal{O}_2\right\},
\ee
from which we can derive the spectral indices and the tensor-to-scalar ratio

\be
n_{\text{t}}=-6\alpha_k^2\left[1+2\alpha_k\left(1+2\sqrt{1+\zeta_k^2}\right)+\mathcal{O}_2\right],
\ee

\be
n_{\text{s}}-1=-4\alpha_k\sqrt{1+\zeta_k^2}\left[1+\alpha _k \left(\frac{-5+4 c_E}{2 \sqrt{1+\zeta _k^2}}-\frac{31}{6} \sqrt{1+\zeta _k^2}+\frac{7+9 \zeta _k^2}{1+\zeta
   _k^2}\right)+\mathcal{O}_2\right],
\ee

\be\label{eq:rimproved}
r=48 \alpha _k^2 \left\{1+2 \alpha _k \left[1+2 \left(c_E-2\right)\sqrt{1+\zeta _k^2} \right]+\mathcal{O}_2\right\},
\ee
where $c_E=\gamma_E+\ln2$, $\gamma_E$ being the Euler-Mascaroni constant.
Finally, for completeness we also show the number of e-folds, although it is not a good quantity for a perturbative expansion~\cite{Anselmi:2020lpp}. From the definition we get
\be\label{eq:Nimproved}
N_k=-\int_{\alpha_k}^{\alpha_f}\frac{\dd \alpha'}{v(\alpha')\beta_{\alpha}(\alpha')}\simeq \int_{\alpha_k}^{\alpha_f}\frac{\dd \alpha'}{2\alpha'^{2}\sqrt{1+\gamma^2/\alpha'^2}}=\frac{1}{2\gamma}\mathrm{arcsinh}\frac{\gamma}{\alpha_k}+\mathcal{O}(\alpha_k^0).
\ee

Since neither $N_k$ nor $\alpha_k$ is an observable, it is convenient to choose one observable, say $\theta_k\equiv 1-n_{\mathrm{s}}(k)$, and express all the others in terms of it. At leading order, for $\gamma^2>0$, we have
\be
r\simeq 3\left(\theta_k^2-16\gamma^2\right),\qquad \mathcal{P}_{\text{s}}\simeq \frac{M^2\kappa^2}{6\pi^2\left(\theta_k^2-16\gamma^2\right)}, \qquad N_k\simeq \frac{1}{2\gamma}\mathrm{arccsch}\left(\sqrt{\frac{\theta_k^2}{16\gamma^2}-1}\right).
\ee

The case of the opposite sign of $\gamma^2$, which is the relevant one for the data, is obtained by the replacement $\gamma^2\rightarrow-\gamma^2$. In the next section we compare these results with those obtained by expanding in $\gamma$ from the beginning, and with the data.

\sect{Comparison with other approaches}

Recently there has been an interest in higher-curvature corrections to the Starobinsky model as a way to reconcile its predictions with the new data from ACT~\cite{Modak:2025bjv,Ketov:2025cqg,Gialamas:2025ofz,Addazi:2025qra,Bezerra-Sobrinho:2025gfg,DelGrosso:2026zbg}. Since ACT finds a slightly larger value of $n_{\mathrm{s}}$, the relevant sign is $\gamma^2<0$. At fixed number of e-folds, a positive $\gamma^2$ lowers $n_{\mathrm{s}}$ with respect to the Starobinsky value, while a negative one raises it. This is not a problem for our calculations. We parametrized the action~\eqref{eq:R3act} with $\gamma^2>0$ only because in this case the exact de Sitter solution~\eqref{eq:desitterH} exists, which allows one to keep the dependence on $\gamma$ exact and expand in $\alpha$ first, while for $\gamma^2<0$ such a solution does not exist. However, since expanding in $\gamma$ first leads to the same resummed functions, the results can be continued to negative $\gamma^2$. Indeed, we verified that they are analytic in $\gamma^2$, so we can change the sign in front of $\gamma^2$ everywhere, i.e.\ $\gamma^2\rightarrow-\gamma^2$ in~\eqref{eq:R3act}, with the only care that the functions of $\zeta_k$ now have branch points at $\zeta_k=\pm1$. 

The number of e-folds follows from the same replacement. Although~\eqref{eq:Nimproved} is written in terms of $\gamma$, it is an even function of $\gamma$ and therefore depends only on $\gamma^2$. Indeed, it can be rewritten as
\be
N_k\simeq\frac{1}{2\alpha_k}\frac{\mathrm{arcsinh}\,\zeta_k}{\zeta_k}=\frac{1}{2\alpha_k}-\frac{\gamma^2}{12\alpha_k^3}+\mathcal{O}(\gamma^4),
\ee
and for the opposite sign of $\gamma^2$ the $\mathrm{arcsinh}$ becomes an $\mathrm{arcsin}$
\be
N_k\simeq\frac{1}{2\alpha_k}\frac{\mathrm{arcsin}\,\zeta_k}{\zeta_k}=\frac{1}{2\alpha_k}+\frac{\gamma^2}{12\alpha_k^3}+\mathcal{O}(\gamma^4).
\ee

The scalar spectral index reads
\be
n_{\mathrm{s}}=1-4\alpha_k\sqrt{1-\zeta_k^2}+\mathcal{O}_2,
\ee
which is real only for $\zeta_k\leq 1$. At first order in $\gamma^2$ it reduces to
\be
n_{\mathrm{s}}\simeq 1-4\alpha_k+\frac{2\gamma^2}{\alpha_k},
\ee
which is the result of recent works, for example~\cite{Gialamas:2025ofz}, where the expansion in $\gamma$ is performed from the very beginning, generating inverse powers of $\alpha$. Without knowing that this formula comes from the expansion of a square root, it could allow values of $\gamma$ that give a complex $n_{\mathrm{s}}$. Imposing $\zeta_k\leq1$ does not significantly change the allowed values of $\gamma$ found in~\cite{Gialamas:2025ofz}.

We also want to highlight that the discrepancy between Starobinsky predictions and the ACT measurements are slightly above the two-sigma range only if we fix the number of e-folds $N_*\equiv N_{k_*}$ between 50 and 60. There is no reason to force it to be in that range, since those numbers are obtained from an estimate and not from a measured quantity (see e.g.~\cite{Gorbunov:2011zzc} for the derivation of this estimate). The correct approach should be the one proposed in~\cite{Piva:2025cqi} and also explained in the previous section, namely to write every quantity in terms of $n_{\mathrm{s}}$. Then the number of e-folds is a prediction of the theory. Starobinsky model predicts roughly $N_*=65$ with the previous Planck 2018 data at central value, while it predicts $N_*\simeq 68-86$ in the one-sigma region of ACT. The estimate of $50-60$ e-folds usually assumes instantaneous reheating. Therefore, we could just allow for larger values of $N_*$, keep Starobinsky model and modify reheating. This is what was done in recent papers ~\cite{Zharov:2025zjg} and~\cite{Drees:2025ngb}. Nevertheless, for the sake of this section, and to compare our results with those of the recent literature, we write the leading-order observables in terms of $N_k$, keeping in mind that we do not necessarily impose a very rigid constraint on the number of e-fold, as it is typically done, for the reasons just explained.

\begin{figure}[t]
    \centering
    \includegraphics[width=0.6\linewidth]{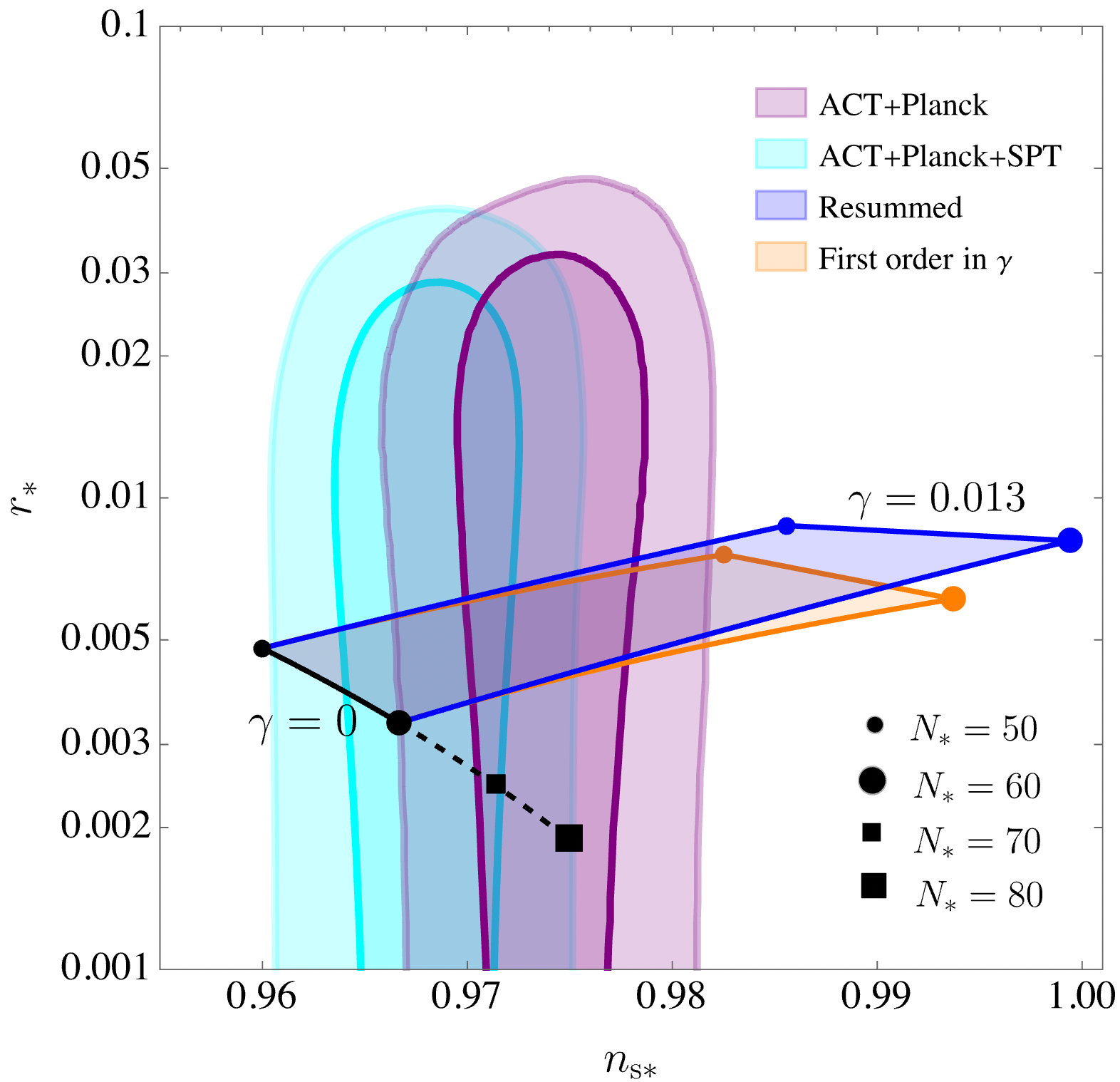}
    \caption{Predictions in the $(n_{\mathrm{s}*},r_*)$ plane. Blue: resummed expansion.
Orange: expansion at first order in $\gamma$. The bands are obtained by
varying $\gamma$ from $0$ to $0.013$ for $N_*=50$--$60$. The black segment is
the Starobinsky model ($\gamma=0$), whose predictions for $N_*=70,80$ are
shown with a dashed line. The purple and cyan contours are the constraints of ACT+Planck~\cite{ACT:2025tim} and ACT+Planck+SPT~\cite{Balkenhol:2025wms}, respectively, in the $1\sigma$ and
$2\sigma$ regions.}
    \label{fig:nsvsr}
\end{figure}

As mentioned above, in order to accommodate a larger value of $n_\mathrm{s}$ we need $\gamma^2<0$. Therefore, we start from the formulas~\eqref{eq:Ptimproved}--\eqref{eq:rimproved} at the leading order for $\gamma^2<0$. From~\eqref{eq:Nimproved} we can derive $\alpha_k$ as a function of $N_k$
\be
\alpha_k=\frac{\gamma}{ \sin\left(2N_k\gamma\right)}, \qquad 4N_k\gamma\leq \pi,
\ee
which we can use to write observable as functions of $N_k$
\be
n_{\mathrm{s}}-1=-\frac{4\gamma}{\tan \left(2N_k\gamma\right)}=-\frac{2}{N_k}\left(1-\frac{4N_k^2\gamma^2}{3}\right)+\mathcal{O}(\gamma^4), 
\ee

\be
r=\frac{48\gamma^2}{\sin^2\left(2N_k\gamma\right)}=\frac{12}{N_k^2}\left(1+\frac{4N_k^2\gamma^2}{3}\right)+\mathcal{O}(\gamma^4),
\ee

\be
\mathcal{P}_{\mathrm{s}}=\frac{M^2\kappa^2}{96\pi^2}\frac{\sin^2\left(2N_k\gamma^2\right)}{\gamma^2}=\frac{M^2N_k^2\kappa^2}{24\pi^2}\left(1-\frac{4N_k^2\gamma^2}{3}\right)+\mathcal{O}(\gamma^4).
\ee

In Fig.~\ref{fig:nsvsr} we show the predictions in the $(n_s,r)$ plane for the
resummed expansion (blue) and for the expansion at first order in $\gamma$
(orange). The black segment is the Starobinsky model ($\gamma=0$), with the
markers indicating $N_*=50$ and $60$, which we take as reference values for a
comparison with the literature. Since $N_*$ is not an observable
and depends on the reheating history, which we do not model, we also show the
Starobinsky predictions for $N_*=70$ and $80$ (dashed line). The colored
bands are obtained by increasing $\gamma$ from $0$ to $0.013$ at fixed $N_*$,
and are bounded by the curves with $N_*=50$ and $N_*=60$. The purple and cyan
contours are the current constraints from ACT+Planck and ACT+Planck+SPT,
respectively. The inclusion of SPT lowers the preferred value of $n_s$~\cite{Balkenhol:2025wms}, so
that the Starobinsky model with $N_*=60$ lies well inside the $1\sigma$ region
of ACT+Planck+SPT, while $N_*=70$ lies inside the $1\sigma$ region of both
data sets and $N_*=80$ is close to the $2\sigma$ boundary of
ACT+Planck+SPT. In the region favored by the data the resummed and the
first-order results are practically indistinguishable, so the expansion at
first order in $\gamma$ is a good approximation as long as
$\gamma/\alpha_k\lesssim1$. They separate for $n_s\gtrsim0.98$, outside the
$2\sigma$ regions of both data sets, where the resummation is needed to
control the corrections.

\begin{figure}[t]
\centering
\includegraphics[width=1\textwidth]{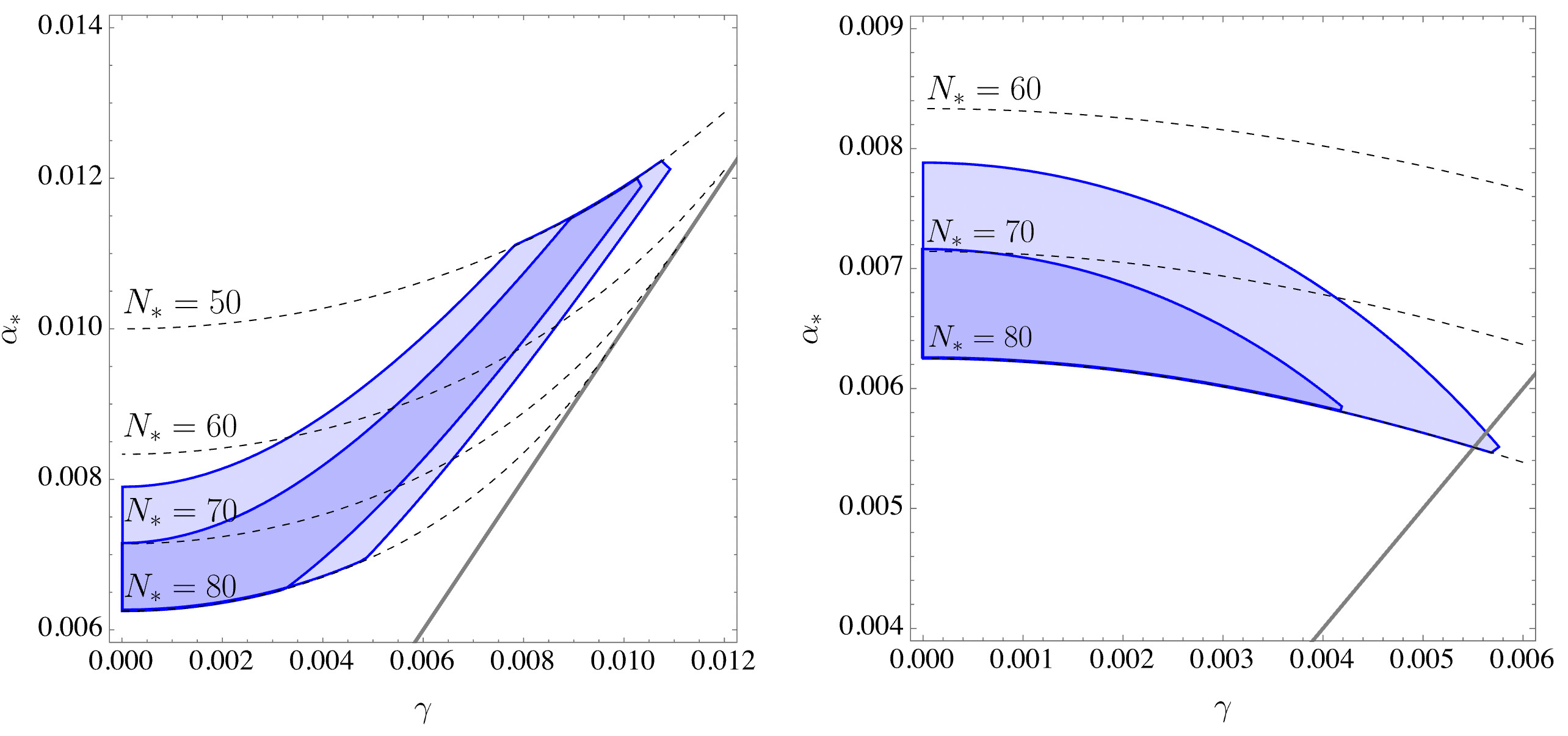}
\caption{Values of $\alpha_*$ and $\gamma$ allowed by ACT+Planck at $1\sigma$ (dark blue) and $2\sigma$ (light blue), for $\gamma^2<0$ (left) and $\gamma^2>0$ (right). The dashed lines are the curves at fixed number of e-folds $N_*=50,60,70,80$ (from top to bottom), and the solid gray line is $\gamma=\alpha_*$, i.e.\ $\zeta=1$. For $\gamma^2<0$ the resummed results are defined only for $\zeta<1$.} 
\label{fig:alphagamma}
\end{figure}

Figure~\ref{fig:alphagamma} shows the values of $\alpha_*\equiv \alpha_{k_*}$ and $\gamma$ allowed by ACT+Planck. For $\gamma^2<0$ (left) the allowed region lies in the domain $\zeta=\gamma/\alpha_*<1$. For $\gamma^2>0$ (right) the $2\sigma$ region extends up to $\zeta\simeq1$ and slightly beyond. There the Taylor expansion in $\zeta$, and hence the expansion at first order in $\gamma$, ceases to converge, while the resummed functions remain well defined. In the rest of the allowed region $\zeta<1$ the two descriptions agree. The number of e-folds lies between $50$ and $80$ and is determined by the point in the $(\gamma,\alpha_*)$ plane, as discussed above.

To quantify how far the EFT description can be trusted, we estimate the scale $\Lambda$ suppressing the $R^3$ term. We regard the $R+R^2$ theory as the base theory, in which $M$ should not be considered as a cutoff, and $R^3$ as the first higher-dimensional operator generated by physics above $\Lambda>M$. With the overall factor $M_{\mathrm{Pl}}^2/2$ common to all terms, operators of dimension $2n$ are expected to appear with coefficients $c_n/\Lambda^{2n-2}$, with $c_n=\mathcal{O}(1)$. Matching with the cubic term of the action, $\gamma^2/(16M^4)\simeq1/\Lambda^4$, gives
\begin{equation}
\Lambda\simeq\frac{2M}{\sqrt{\gamma}}\,.
\end{equation}
For $N_*=60$, the central value of $n_s$ of ACT+Planck corresponds to $\gamma\simeq6.7\times10^{-3}$, hence
\be
\Lambda\simeq24\,M\simeq3\times10^{-4}M_{\mathrm{Pl}}
\ee
for $M\simeq1.2\times10^{-5}M_{\mathrm{Pl}}$. The parameter $M$ is fixed by the amplitude of scalar perturbations and depends only mildly on $\gamma$. It changes by about $10\%$ for $\gamma\lesssim0.007$, so that the value used above for the estimate of $\Lambda$ is accurate for our purposes. During inflation $H^2\simeq NM^2/3$ and $R\simeq-4NM^2$, so that
\begin{equation}
\frac{H}{\Lambda}=\frac{1}{2}\sqrt{\frac{N_*\gamma}{3}}\simeq0.2, \qquad -\frac{R}{\Lambda^2}=N_*\gamma\simeq0.4\,.
\end{equation}
Therefore, the explanation of the ACT shift through an $R^3$ term with a tree-level coefficient places us at the limit of validity of the effective theory. 

If the higher-dimensional operators are instead generated by quantum corrections, their coefficients are further suppressed by a loop factor, $c_n\sim1/16\pi^2$, which gives a lower scale
\be
\Lambda\simeq \frac{M}{\sqrt{\pi\gamma}}\simeq7\,M\simeq8\times10^{-5}M_{\mathrm{Pl}},
\ee
for which 
\be
\frac{H}{\Lambda}\simeq0.7, \qquad -\frac{R}{\Lambda^2}\simeq5,
\ee
so that the EFT would hardly be under control.

Evaluated on the inflationary background, the $R^3$ term itself is only a $\sim0.4\%$ correction to the $R^2$ term in the action, but the observables are sensitive to it because of the enhancement discussed above.

Operators of higher dimension, such as $R^4$, are not suppressed a priori. If their coefficients are natural in units of $\Lambda$, $\delta/M^6\sim1/\Lambda^6$, they contribute with respect to $R^3$ by a factor $|R|/\Lambda^2\simeq0.4$, since the relevant curvature scale is that of the inflationary solution. The truncation at $R^3$ cannot then be justified by power counting alone, and requires a hierarchy among the coefficients. The value of $\gamma$ should then be interpreted as an effective parameter, summarizing unknown ultraviolet physics at a scale not far above $H$. This raises the question of whether a UV-complete theory can generate such terms with the required coefficients. In the next section we address it in the case of renormalizable quantum gravity.

\sect{Comment on renormalizable gravity}

Higher-curvature terms are expected to originate from UV-complete models, for example renormalizable and superrenormalizable theories of quantum gravity~\cite{Stelle:1976gc,Modesto:2016ofr,Calcagni:2022shb, Modesto:2026utk}. As a concrete example, we consider the theory
\be\label{eq:QGact}
S_{\mathrm{QG}}(g)=-\frac{1}{2\kappa^2}\int\sqrt{-g}\left[R+2\Lambda_C-\frac{1}{6m_{\phi}^2}R^2+\frac{1}{2m_{\chi}^2}C_{\mu\nu\rho\sigma}C^{\mu\nu\rho\sigma}\right]
\ee
equipped with the fakeon prescription~\cite{Anselmi:2017yux,Anselmi:2017ygm,Anselmi:2018ibi,Anselmi:2018tmf} to make it also unitary. Such a theory is renormalizable and contains three scales (neglecting the cosmological constant): the Planck mass $M_{\mathrm{Pl}}$, the scalaron mass $m_{\phi}$ and the spin-2 fakeon mass $m_{\chi}$. We assume the scalaron to drive inflation, so that $m_\phi$ corresponds to the mass $M$ of the previous sections. The inflationary predictions for scalar perturbations coincide with those of the Starobinsky model up to the NNL order, where $m_\chi$ first appears, while the tensor-to-scalar ratio is modified already at leading order~\cite{Anselmi:2020lpp}. Indeed, the theory produces values of $r$ constrained in the window $0.37\lesssim 1000r\lesssim 3.41$, roughly one order of magnitude wide~\cite{Anselmi:2020lpp}, which could be tested by the next-generation experiments for detecting the polarization of the CMB, such as LiteBIRD~\cite{Hazumi:2019lys}. The right extreme corresponds to $m_{\chi}\rightarrow \infty$, i.e.\ the Starobinsky model, while the other extreme follows from the consistency of the fakeon approach, giving the bound $m_{\chi}> m_{\phi}/4$ (see~\cite{Anselmi:2020lpp} and~\cite{Piva:2023bcf} for details).

The sensitivity of the observables to higher-curvature terms, even within the regime of validity of the EFT, raises the question of how robust the predictions of Starobinsky inflation are. As an effective theory, it cannot exclude new physics at scales between $H$ and the scale $\Lambda$ estimated in the previous section, which would generate higher-curvature operators with unknown coefficients and might shift $n_s$ by an amount comparable to the current uncertainty. Instead, the effective action of the theory~\eqref{eq:QGact} contains local divergent terms up to dimension four, as well as nonlocal finite terms, while local higher-dimensional operators arise only when we expand in inverse powers of some mass, with coefficients that are fixed by the parameters of the theory. During inflation we can only expand in $H/m_{\chi}$, with $H\simeq4.5\,m_\phi$. If $m_{\chi}$ is comparable to $m_\phi$ we cannot perform such an expansion, and only nonlocal form factors appear. On the other hand, if $m_{\chi}\gg m_{\phi}$ local higher-curvature terms are generated by the expansion. However, in both cases these terms do not introduce new free parameters. They are generated by loops and suppressed by the loop factor $1/16\pi^2$ and by the dimensionless couplings of the theory, of order $m_\phi^2/M_{\mathrm{Pl}}^2\sim10^{-10}$ and $m_\chi^2/M_{\mathrm{Pl}}^2$ (small as long as $m_\chi\ll M_{\mathrm{Pl}}$), i.e.\ $m_\phi^2/(16\pi^2M_{\mathrm{Pl}}^2)\sim10^{-12}$ and $m_\chi^2/(16\pi^2M_{\mathrm{Pl}}^2)$ overall. A rough power counting indicates that their effect on $n_s$ is expected to lie many orders of magnitude below the value of $\gamma\sim10^{-2}$ needed to shift it appreciably, so that we are in the regime $\gamma\ll\alpha$, where the expanded and the resummed results are compatible with the Starobinsky ones. Therefore, the predictions are those of the Starobinsky model supplemented by the fakeon contribution, mainly affecting $r$, and they are stable under quantum corrections. The price is that, taking the ACT+Planck central value $n_s\simeq0.974$, a number of e-folds $N_*\simeq78$ is required, constraining the post-inflationary history. The inclusion of SPT data lowers the central value to $n_s\simeq0.968$~\cite{Balkenhol:2025wms}, corresponding to $N\simeq63$, so that this requirement is considerably relaxed.

\sect{Conclusions}

We have studied Starobinsky inflation supplemented by higher-curvature operators, focusing on the term $\gamma^2R^3/(16M^4)$. When the observables are expanded in the slow-roll parameter $\alpha$ at fixed $\gamma$, or in $\gamma$ at fixed $\alpha$, the coefficients of the expansions contain inverse powers of $\gamma$ or of $\alpha$, respectively. Therefore, it is not clear a priori how to organize the expansions when $\gamma$ and $\alpha$ are comparable. Using the running of the slow-roll parameter and the analogy with the RG, we have shown that this problem is solved by means of a resummation. The terms with inverse powers are collected into functions of the single variable $\zeta=\gamma/\alpha$, and the expansion is organized in powers of $\alpha$ at fixed $\zeta$. In this way each order of the new improved slow-roll expansion is known as an exact function of $\zeta$, and no inverse powers of $\gamma$ or $\alpha$ are left. As a consequence also the power spectra, the spectral indices and the tensor-to-scalar ratio can be written in the improved expansion. We have also shown that the expansions in $\alpha/\gamma$ and in $\gamma/\alpha$ lead to the same resummed functions.  Therefore, the resummed results are independent of the order in which the double expansion is performed.

The comparison with the literature clarifies the role of the first-order expressions for $n_s$ and $r$, which are the Taylor expansion of our results in $\gamma$, truncated at the first term in $\zeta$. In the region favored by ACT+Planck and by ACT+Planck+SPT the two descriptions are practically indistinguishable, because the data select $\zeta<1$. They separate only for $n_s\gtrsim0.98$, where the resummation is needed.

Moreover, the size of $\gamma$ needed to move $n_s$ corresponds to an EFT cutoff $\Lambda=2M/\sqrt\gamma\sim10^{-4}M_{\rm Pl}$, only a factor of a few above the Hubble scale during inflation. This estimate is lower if the operators are assumed to be generated at loop level. In this regime operators of higher dimension, such as $R^4$, are not suppressed by power counting alone, and the value of $\gamma$ must be regarded as an effective parameter summarizing unknown ultraviolet physics.

Finally, we have commented on the stability of the predictions of the Starobinsky model and of renormalizable quantum gravity with fakeons. In the latter case no free higher-dimensional local operators are present and the scalar spectrum coincides with the Starobinsky one up to NNL order, while the main departure is in the tensor-to-scalar ratio. The crucial difference is that, thanks to its renormalizability, the inflationary predictions of quantum gravity are stable under quantum corrections, while in the Starobinsky model, as an effective theory, they are not.

\vskip.3 truecm \noindent {\large \textbf{Acknowledgments}}

\vskip .5 truecm

We are grateful to D. Anselmi, P. Baratella and A. Melis for useful discussions.
\appendix
\renewcommand{\thesection}{\Alph{section}} \renewcommand{\theequation}{%
\thesection.\arabic{equation}} \setcounter{section}{0}

\bibliographystyle{JHEP} 
\bibliography{mybiblio}

\end{document}